\documentstyle[twocolumn,prl,aps,psfig]{revtex}        

\newcounter{FAQ}
\begin{document}
\twocolumn[\hsize\textwidth\columnwidth\hsize\csname@twocolumnfalse\endcsname
\draft

\title{The Mott-Anderson transition in the disordered one-dimensional
  Hubbard model}

\author{
Ramesh V. Pai,${^{1,2}}$
Alexander Punnoose${^{3}}$ and
Rudolf A.\ R\"{o}mer${^{4}}$\\
${^1}$Department of Physics,
Goa University,
Goa, 403 206, India\\
${^{2}}$Jawaharlal Nehru Center for Advanced Scientific Research,
Jakkur, Bangalore, 560 064, India\\
${^{3}}$Department of Physics,
Indian Institute of Science,
Bangalore, 560 012, India\\
${^{4}}$Institut f\"{u}r Physik,
Technische Universit\"{a}t Chemnitz,
D-09107 Chemnitz, Germany
}

\date{Version: April 2, 1997; printed \today}
\maketitle

\begin{abstract}
  We use the density matrix renormalization group to study the quantum
  transitions that occur in the half-filled one-dimensional fermionic
  Hubbard model with onsite potential disorder. We find a transition
  from the gapped Mott phase with algebraic spin correlations to a
  gapless spin-disordered phase beyond a critical strength of the
  disorder $\Delta_c \approx U/2$. Both the transitions in the charge
  and spin sectors are shown to be coincident. We also establish the
  finite-size corrections to the charge gap and the spin-spin
  correlation length in the presence of disorder and using a
  finite-size-scaling analysis we obtain the zero temperature phase
  diagram of the various quantum phase transitions that occur in the
  disorder-interaction plane.
\end{abstract}

]

%
%


Electronic systems can undergo quantum phase transitions from metallic
to insulating behavior as a function of either the interaction
strength or the degree of disorder or both \cite{tvr}. A clean system
at certain commensurate fillings may develop a gap in the energy
spectrum as the strength of the repulsion is increased and turn into a
Mott insulator \cite{mott}.  On the other hand, a system of
noninteracting electrons can undergo a transition from metal to
insulator as the degree of randomness is increased and turn into an
Anderson localized insulator \cite{anderson}. The interplay of
electron-electron interaction and disorder raises interesting
possibilities of a new type of transition, distinct from the clean
correlation induced Mott transition or
the disorder induced Anderson transition.


The repulsive Hubbard model in one dimension (1D) is probably the
simplest model which shows a Mott transition at half-filling for
arbitrary values of the interaction strength $U>0$ \cite{lieb}. One of
the most attractive features of this model is that the Mott transition is
unaccompanied by a spin-density-wave gap as {\em e.g.}\ happens in the 2D
Hubbard model due to the existence of magnetic long-range order (MLRO)
\cite{mlro}. Instead the 1D model shows algebraically decaying spin
correlations as the maximal remnant of MLRO in 1D. The properties of the
clean 1D Hubbard model are well established: The dependence of the Mott
gap on the interaction parameter \cite{lieb}, the asymptotic behavior
and the critical exponents of various correlation functions \cite{frahm}
have all been computed using Bethe Ansatz and the finite-size-scaling
approach of conformal quantum field theory \cite{bzp}. All this makes
this model particularly attractive to study the effect of disorder on
the Mott state.


Recent studies of the half-filled disordered 1D Hubbard model have
proceeded numerically using the Quantum Monte Carlo (QMC) method
\cite{scalapino} and analytically using bosonization and the
renormalization group method \cite{kawakami}. The QMC results give a
very accurate description of the finite temperature properties of the
system. However, the low temperature properties can only be inferred
by extrapolating the finite temperature data and going to larger
system sizes. On the other hand, bosonization methods using {\em
  perturbative} renormalization group techniques only give an
indication of the various plausible fixed points in parameter space.
Computing correlation functions at the strong coupling fixed points is
then impossible as the coupling constants are driven away from their
weak coupling values.  Nevertheless, the results of both approaches
indicate that at half-filling a {\em finite} amount of potential
disorder is needed to cause a transition from Mott (gapped) to
Anderson (gapless) insulating behavior. This is in qualitative
agreement with arguments put forward by Ma \cite{ma}.

In the present Letter, we have studied the ground state properties of
the disordered 1D Hubbard model at zero temperature with the help of
the density-matrix renormalization group (DMRG) \cite{white}. This
method has been previously shown to be highly successful for 1D
quantum systems \cite{spinchains,boseramesh} and may be seen as a
numerical variational-wave-function approach \cite{stellan}.
After introducing the parameters of our DMRG, we show that in the
clean case, the previously known results can be reproduced numerically
reliably. We then consider finite disorder and show that the charge
gap $G^c$ remains open for small disorder up to a critical disorder
strength $\Delta_c \approx U/2$. The transition in the spin sector is
seen by studying the behavior of the spin-spin-correlation function
$\langle S^-(r) S^+(0)\rangle$. For $\Delta < \Delta_c$, the power-law
remnant of the MLRO persists, whereas for $\Delta > \Delta_c$, the
spin-spin correlation indicates the emergence of a spin-disordered
phase.

%
%


The Hubbard Hamiltonian with additional potential disorder on a chain
of $L$ sites is given as
\begin{equation}
\label{eq-hamiltonian}
H= -t
\sum_{\stackrel{\scriptstyle{x=1}}{\sigma=\uparrow,\downarrow}}^{L}
        ({c}_{x+1\sigma}^{\dagger}{c}_{x\sigma} + h.c.) +
U \sum_{x=1}^{L} {n}_{x\uparrow}{n}_{x\downarrow} +
\sum_{x=1}^{L} \mu_x{n}_{x},
\end{equation}
where $-t$ is the hopping amplitude between nearest-neighbor,
${c}_{x\sigma}^{\dagger}({c}_{x\sigma})$ the fermion creation
(annihilation) operator at site $x$ with spin $\sigma$, ${n}_
{x\sigma}={c}_{x\sigma}^{\dagger}{c}_{x\sigma}$ the number operator,
${n}_{x}={n}_{x\uparrow}+{n}_{x\downarrow}$ and $U$ is the onsite
repulsive energy. The onsite chemical potential $\mu_{x}$ is a random
number which we take to be uniformly distributed between $\pm\Delta$
such that $\Delta=0$ corresponds to the clean case. We work at
half-filling, {\em e.g.}\, $N=\langle\sum_{x=1}^{L}{n}_x\rangle=L$ and the
energy scale is set by choosing $t=1$.


We follow the standard open chain DMRG algorithm of White
\cite{white}. In each iteration, we diagonalize the Hamiltonian matrix
of a super-block denoted by $B^{l}_{L/2-1}\bullet \bullet\ 
B^{r}_{L/2-1}$ of $L$ sites and obtain the energy and the wave
function $|\psi_{0L}\rangle$ of the ground state \cite{boseramesh}. Here
$\bullet$ represents a single site.  Using $|\psi_{0L}\rangle$ as the
target state \cite{white} we compute the reduced density matrix
$\rho^{l}$ of the left sub-block $B^{l}_{{L/2}} \equiv B^{l}_{{L/2-1}}
\bullet$ of size $L/2\/$.  Diagonalizing $\rho^{l}$ we obtain its
eigenstates and the eigenvalues. These eigenstates with the highest
eigenvalues are the most probable states of the left sub-block when
the super-block is in the state $|\psi_{0L}\rangle$ and so can be used for
truncation.  Keeping $M$ eigenstates corresponding to the largest $M$
eigenvalues of $\rho^{l}$ as the new basis of the left sub-block, we
transform the Hamiltonian and all further operators into this new
basis. Since every single $\bullet$ has $4$ states, we thus truncate
the original $4M$ states of $B^{l}_{L/2}$ to only $M$ states. Usually
we have used $M=128$. Because the disorder destroys translational
symmetry, the left and right sub-blocks are non-identical. Hence a
similar procedure needs to be followed for the right sub-block
$B^{r}_{L/2}$.  The above steps are then repeated for the new
super-block $B^{l}_{L/2}\bullet \bullet\ B^{r}_{L/2}$ with $L+2$ sites
and thus the system increases by two sites at each iteration.

Let $E_{0}(L)$ denote the ground state energy of the Hubbard chain of
length $L$ with $N_{\uparrow}=N_{\downarrow}=L/2$. In order to compute
the charge gap $G_{L}^{c}$ for a system of finite length $L$, we
repeat the DMRG steps with $N_{\uparrow}=L/2+1$ and
$N_{\downarrow}=L/2$ and also with $N_{\uparrow}=L/2-1$ and
$N_{\downarrow}=L/2$. We denote these two ground state energies by
$E_{1}(L)$ and $E_{-1}(L)$, respectively. The charge gap is then
defined as the discontinuity of the chemical potential at half-filling
\cite{lieb}, {\em i.e.}, $G_{L}^{c}=E_{1}(L)+E_{-1}(L)-2E_{0}(L)$. In the
presence of disorder $G_{L}^{c}$ is computed for at least $10$
different disorder realizations and then averaged over all such
realizations.
We also compute the spin-spin-correlation function $\Gamma_{L}^{s}(r)
= \langle \psi_{0L}| S^{-}(r)S^{+}(0) |\psi_{0L}\rangle$ and the
second moment of the staggered antiferromagnetic (AFM) correlation
function $(\xi_{L}^{s})^{2}= \sum_r r^2(-1)^r\Gamma_{L}^{s}(r)$. Here,
$S^{+}(r)={c}_{r\uparrow}^{\dagger}{c}_{r\downarrow}$. In the presence
of disorder, the correlation functions are first averaged over the
disorder realizations and then the correlation length $\xi_{L}^{s}$ of
the averaged staggered correlation function is found.



\begin{figure}
\centerline{\psfig{figure=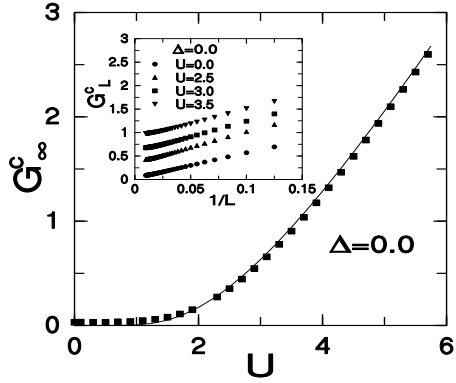,width=7.0cm,height=6.0cm}}
 \caption{Filled squares (\protect\rule[0.1ex]{.9ex}{.9ex}) show the charge gap
   $G_{\infty}^{c}$ as a function of the interaction strength $U$.
   The exact result \protect\cite{lieb} is shown by the solid line.
   Inset: $G_{L}^{c}$ vs $1/L$ for $U=0.0,2.5,3.0$ and $3.5$.  The
   value of the intercept gives $G_{\infty}^{c}$.}
\label{fig-clean-ginfty}
\end{figure}

\begin{figure}
\centerline{\psfig{figure=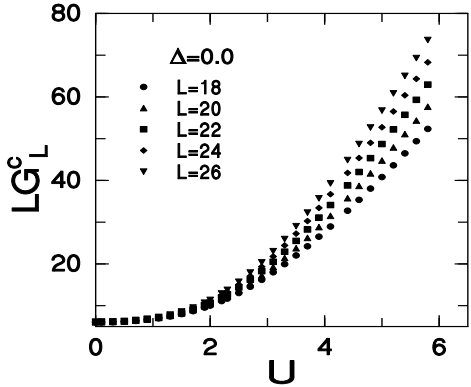,width=7.0cm,height=6.0cm}}
\caption{$LG^{c}_{L}$ as a function of the interaction strength $U$
  showing the coalescence of curves for different $L$ at $U=0$. This
  indicates that the critical value of the interaction strength at
  which the Mott transition occurs is $U=0$.  }
\label{fig-clean-LG}
\end{figure}

{\em The clean case $\Delta=0$}:
It has been shown in Ref.\ \cite{lieb} that the half-filled repulsive 1D
Hubbard model exhibits a charge gap for all non-zero values of the
interaction strength. We find that the functional dependence of the
thermodynamic value of the charge gap $G_{\infty}^{c}$ on the
interaction strength $U$ as obtained by DMRG (Fig.\
\ref{fig-clean-ginfty}) is in excellent agreement with the exact
solution computed in Ref.\ \cite{lieb},
\begin{equation}
G^{c}=U-4+8\int_{0}^{\infty} dw {J_{1}(w)\over w(1+\exp{(wU/2)})}
\label{eq-lieb}
\end{equation}
with $J_{1}(w)$ a Bessel function. We note that the DMRG algorithm using
open boundary conditions gives consistent results for the extrapolated
charge gap $G_{\infty}^{c}$ with that of Eq.\ \ref{eq-lieb} which has been
derived using periodic boundary conditions.
Nevertheless, we now employ a finite-size-scaling (FSS) analysis of
the charge gap $G_{L}^{c}$ in order to remove any finite-size effects
that could arise in the extrapolation. In Fig.\ 
{\ref{fig-clean-ginfty}} (inset) we show that the leading order
finite-size corrections to $G_{\infty}^{c}$ fall off as $1/L$, {\em
  i.e.} $G_{L}^{c}(U) = G_{\infty}^{c}(U)+g(U)/L $.  Having determined
the explicit scale dependence of $G_{L}^{c}$, a plot of $LG_{L}^{c}$
versus $U$ as in Fig.\ \ref{fig-clean-LG} shows curves for different
$L$ coalescing as the charge gap vanishes. This finite-size behavior
allows a numerically accurate determination of the critical value of
the interaction strength and we find in accordance with the result of
Ref.\ \cite{lieb} that the Mott transition occurs at $U=0$.

\begin{figure}
\centerline{\psfig{figure=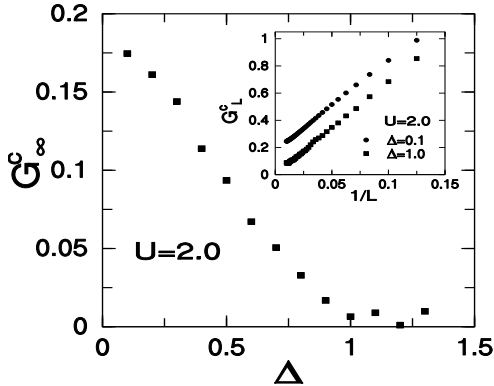,width=7.0cm,height=6.0cm}}
\caption{Filled squares (\protect\rule[0.1ex]{.9ex}{.9ex}) show the charge gap
  $G_{\infty}^{c}$ as a function of the disorder $\Delta$ for $U=2$.
  Inset: $G_{L}^{c}$ vs $1/L$ for $\Delta=0.1\mbox{ and }1.0$.  The
  value of the intercept gives $G_{\infty}^{c}$.  }
\label{fig-d20-ginfty}
\end{figure}
\begin{figure}
\centerline{\psfig{figure=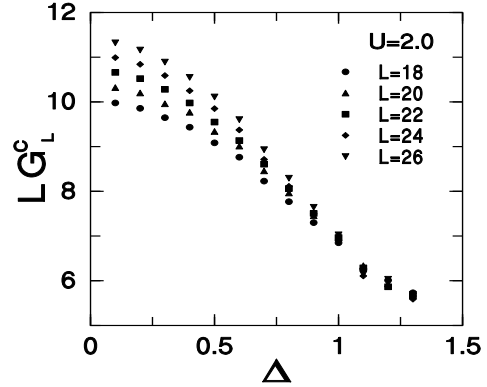,width=7.0cm,height=6.0cm}}
\caption{$LG^{c}_{L}$ as a function of the disorder $\Delta$ for $U=2$
  showing the coalescence of the curves for different $L$ at
  $\Delta=1.0$. This indicates the transition from a gapped Mott
  insulator phase to a gapless phase.  }
\label{fig-d20-LG}
\end{figure}

{\em The disordered case $\Delta \neq 0$}:
Quantum Monte Carlo \cite{scalapino} and bosonization \cite{kawakami}
studies have predicted the existence of a critical disorder $\Delta_c
> 0$ beyond which the Mott gap vanishes. In the limit of large $U$,
this can be motivated \cite{shankar} by potential energy
considerations: any rearrangement of the one particle per site
configuration in the half-filled Hubbard model would necessarily cost
an energy $U$ due to double occupancy and would gain in the local site
potential energy a maximum of $2\Delta$. Hence, in order for it to be
feasible for the electrons to take advantage of the random site
energies we should have $\Delta_c \geq U/2$ in the case of bounded
disorder.
In Fig.\ \ref{fig-d20-ginfty} we show $G_{\infty}^{c}$ as a
function of the disorder strength $\Delta$ for $U=2$. We see that the
system undergoes a transition from a gapped Mott insulator phase to a
gapless phase for $\Delta_{c} \approx 1 = U/2$. Fig.\ 
\ref{fig-d20-ginfty} (inset) shows the finite-size corrections to
$G_{\infty}^{c}$ in the presence of two representative weak
($\Delta=0.1$) and strong ($\Delta=1.0$) disorder strengths for $U=2$.
We note that the corrections continue to fall off as $1/L$ even in the
presence of disorder. The FSS plot of $LG_{L}^{c}$ versus $\Delta$
(Fig.\ \ref{fig-d20-LG}) shows that curves for different $L$ coalesce
at $\Delta_{c}=1.0 \pm 0.1$ again indicating the transition from a
gapped to a gapless phase as the strength of the disorder is
increased. We have also done the same analysis for $U=3$ and $5$ where
a well developed Mott gap exists and where the DMRG is more
stable\cite{white}. Our results indicate that the transition into the
gapless phase takes place at $\Delta_c \approx U/2$.

\begin{figure}
\centerline{\psfig{figure=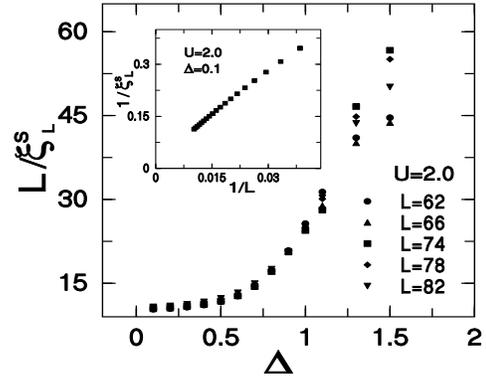,width=7.0cm,height=6.0cm}}
 \caption{$L/{\xi_{L}^{s}}$ as a function of the disorder $\Delta$
   for $U=2$ showing the coalescence of the curves for different $L$
   at $\Delta\approx 0.9-1.0$.  This indicates the transition from a
   gapped Mott insulator phase to a spin-disordered gapless phase.
   Inset: $1/{\xi_{L}^{s}}$ vs $1/L$ for $\Delta=0.1$.  }
\label{fig-d20-LXi}
\end{figure}

{\em Spin-spin-correlation function}:
For $\Delta=0$, the half-filled 1D Hubbard model shows algebraically
decaying antiferromagnetic (AFM) correlations as computed by the
methods of conformal field theory
and bosonization \cite{frahm},
\begin{equation}
  \Gamma^{s}(r) \approx A_{1}{\cos(\pi r +\phi_{1})\over r}+ {\cal O}(1/r^2),
\label{eq-frahm}
\end{equation}
When the spin-spin-correlation function decays as a power-law, as in
Eq.\ \ref{eq-frahm}, the AFM correlation length $\xi^{s}_{\infty}$
diverges in the thermodynamic limit. However, in the absence of
long-ranged correlations, $\xi_{\infty}^{s}$ remains finite. Fig.\ 
\ref{fig-d20-LXi} (inset) shows that for $\Delta=0.1$ the finite-size
correction to $1/\xi_{\infty}^{s}$ continues to fall off as $1/L$ such
that
$1/\xi_{L}^{s}(\Delta)=1/\xi_{\infty}^{s}(\Delta)+\zeta(\Delta)/L$ and
we can also apply the previous FSS analysis to the correlation length.
In Fig.\ \ref{fig-d20-LXi} we plot $L/\xi_{L}^{s}$ as a function of
$\Delta$. The data for different $L$ coalesce until the disorder
$\Delta_{s} \approx 0.9 \pm 0.1$. Thus for $U=2$ the staggered
$\Gamma^{s}(r)$ continues to fall off with a power-law up to this
critical disorder $\Delta_{s}$. For larger $\Delta$ beyond $\Delta_s$,
the values of $L/\xi_{L}^{s}$ do not coalesce any more.  This
indicates the transition from power-law correlations into a
short-ranged spin-disordered phase with a finite correlation length.
Again, the same analysis of the spin-spin-correlation function for
$U=3$ and $5$ confirms that $\Delta_s \approx U/2$.

\begin{figure}
\centerline{\psfig{figure=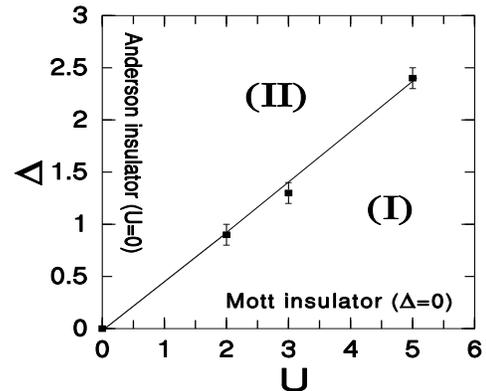,width=7.0cm,height=6.0cm}}
 \caption{The phase diagram of the disordered 1D Hubbard
   Hamiltonian (\ref{eq-hamiltonian}) at half-filling showing the Mott
   insulator with algebraic spin correlations (I) and the gapless spin
   disordered phase (II). The phase boundary has been drawn through
   the computed points (filled squares
   (\protect\rule[0.1ex]{.9ex}{.9ex}) with error bars) and is close to
   $\Delta= U/2$. }
\label{fig-phase}
\end{figure}

Thus the two transitions seen above viz.\ (I) gapped Mott insulator
phase to a gapless phase at $\Delta_c$ and (II) long-ranged AFM phase
to a short-ranged spin-disordered phase at $\Delta_s$, are coincident,
and $\Delta_c = \Delta_s$ within the numerical accuracy. Beyond the
critical disorder $\Delta_c \approx U/2$ the gapped Mott insulator
with power-law AFM correlations goes over to a gapless short-ranged
spin-disordered phase. This is the main result of our work and allows
us to show in Fig.\ \ref{fig-phase} the phase diagram of the zero
temperature quantum transitions that occur in the 1D disordered
half-filled Hubbard model.


In summary, the DMRG allows us to analyze the various quantum
transitions that occur in the half-filled 1D Hubbard model in the
presence of onsite disorder.  For $\Delta=0$ we find that the
finite-size corrections to the thermodynamic value of the charge gap
$G_{\infty}^{c}$ scale as $1/L$ (Fig. \ref{fig-clean-ginfty} (inset)).
The functional dependence of $G_{\infty}^{c}(U)$ on the interaction
strength (Fig.\ \ref{fig-clean-ginfty}) is in excellent agreement with
the exact solution computed in Ref.\ \cite{lieb} in support of our
DMRG approach. We have further shown that for $\Delta\neq 0$, the
finite size corrections to $G_{\infty}^{c}(\Delta)$ continue to fall
off as $1/L$. For small $\Delta < U/2$, the Mott gap is shown to
survive.  Curves of $LG_{L}^{c}$ versus $\Delta$ for different $L$
(Fig.\ \ref{fig-d20-LG}) come together at $\Delta_{c}$ ($\approx 1.0$)
and coalesce after that indicating a transition from the gapped Mott
insulator phase to a gapless phase. A finite-size analysis of the
staggered AFM correlation length $\xi_{L}^{s}$ shows that the values
of $L/{\xi_{L}^{s}}$ for various disorders $\Delta$ (Fig.
\ref{fig-d20-LXi}) also coalesce at $\Delta_s\approx 1.0 \pm 0.1$.
These transitions in the charge and the spin sector are shown to be
coincident $\Delta_c = \Delta_s$. Thus in the gapped Mott phase
$\Gamma^{s}(r)$ decays algebraically and is short ranged in the
disordered gapless phase. We obtain the ($U$, $\Delta$) phase diagram
(Fig.\ \ref{fig-phase}) showing the phase boundary which separates the
gapped Mott insulator phase with algebraic spin correlations and the
spin-disordered gapless phase.

\acknowledgments 

We sincerely thank S.\ Ramashesha for useful discussions on the DMRG
technique and the HRZ (TUCZ) and SERC (IISc) for the computing
facilities.  This work has been supported by the Deutsche
Forschungsgemeinschaft (Sfb 393).

%
%

%
%

\vspace{-1cm}

\vspace{-1cm}

\vspace{-1cm}

\vspace{-1cm}

\vspace{-1cm}

\vspace{-1cm}

\end{document}